\documentclass[twocolumn,superscriptaddress,showpacs,apl]{revtex4}
\usepackage[dvips]{graphicx}
\usepackage{amssymb,amsmath,amsfonts}

\begin{document}
    \title{Terahertz metamaterials on free-standing highly-flexible polyimide substrates }
    \date{\today}\received{}
    \author{Hu Tao}
     \affiliation{Boston University, Department of Manufacturing Engineering, 15 Saint Mary's Street, Brookline, Massachusetts, 02446}
    \author{A.~C.~Strikwerda}
    \affiliation{Boston University, Department of Physics, 590 Commonwealth Ave, Boston, Massachusetts, 02215}
    \author{K. Fan}
     \affiliation{Boston University, Department of Manufacturing Engineering, 15 Saint Mary's Street, Brookline, Massachusetts, 02446}
    \author{C. M. Bingham}
    \affiliation{Boston College, Department of Physics, 140 Commonwealth Ave., Chestnut Hill, MA 02467}
    \author{W. J. Padilla}
    \affiliation{Boston College, Department of Physics, 140 Commonwealth Ave., Chestnut Hill, MA 02467}
    \author{Xin Zhang}
     \email{xinz@bu.edu}
    \affiliation{Boston University, Department of Manufacturing Engineering, 15 Saint Mary's Street, Brookline, Massachusetts, 02446}
    \author{R.~D.~Averitt}
     \email{raveritt@physics.bu.edu}
    \affiliation{Boston University, Department of Physics, 590 Commonwealth Ave, Boston, Massachusetts, 02215}

\begin{abstract}
We have fabricated resonant terahertz metamaterials on free standing
polyimide substrates. The low-loss polyimide substrates can be as
thin as 5.5 $\mu$m yielding robust large-area metamaterials which
are easily wrapped into cylinders with a radius of a few
millimeters. Our results provide a path forward for creating
multi-layer non-planar metamaterials at terahertz frequencies.
\end{abstract}

\maketitle


The advent of metamaterial composites has given rise to numerous
electromagnetic functionalities previously unimagined. This includes
negative refractive index, superlensing, cloaking, and quite
generally, the fabrication of metamaterials which have been designed
using coordinate transformation approaches
\cite{smith00,shelby01,pendry00,schurig06a,pendry06}. Many of these
ideas were initially implemented at microwave frequencies where
fabrication of multilayer composites has become increasingly
sophisticated during the past several years. This has resulted in
dramatically reduced times from conceptualization and
electromagnetic simulation to, ultimately, fabrication and
characterization. However, the fabrication of subwavelength unit
cells becomes increasingly challenging in moving from the microwave
to visible region of the electromagnetic spectrum though important
progress has been made \cite{pendry99,yen04,schurig06,dolling07}. To
date, the majority of this work has been on planar composites. At
terahertz (THz) frequencies and above, creating multiple unit cell
structures in the direction of propagation and taking full advantage
of coordinate transformation MM design to realize non-planar MM
composites requires the development of new fabrication strategies.

The far-infrared, or terahertz, is a promising region to investigate
novel approaches to metamaterials fabrication. First, the unit cells
are on the order of tens of microns which is amenable to novel
microfabrication approaches. Second, and perhaps of greater
importance, there is a strong technological impetus to create
sources, detectors, and components at terahertz frequencies to
realize the unique potential of THz radiation
\cite{williams06,tonouchi07,zhang02,nishizawa05}. Metamaterials are
expected to play an important role in this regard as evidenced by
recent demonstrations of MM-based modulators and frequency tunable
filters \cite{padilla06,chen06,chen08}. An important step in the
progression of functional THz MM composites is the fabrication of
multilayer structures. For example, a strongly resonant THz MM
absorber consisting of two layers spaced by approximately six
microns \cite{tao08} was recently demonstrated (see also
\cite{landy07} for the microwave perfect absorber). The entire
structure, however, was on top of a thick GaAs supporting substrate.
There has also been other work at THz frequencies using polymer
spin-coating based techniques to fabricate metamaterials, but the
focus of this work was not on ultrathin flexible substrates
\cite{katsarakis2005,paul2008}.

\begin{figure}[h]
\includegraphics[scale=0.8]{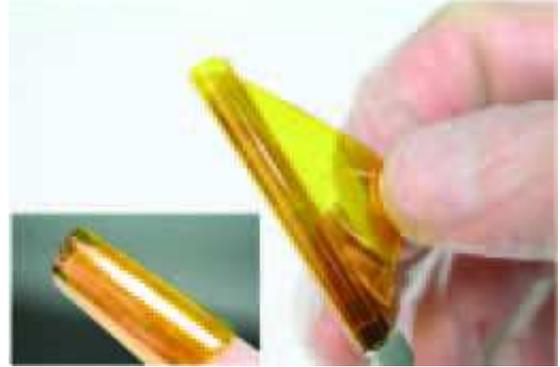}
\caption{\label{fig:Fig1} (Color) Photographs of the free standing
electric metamaterials fabricated on polyimide substrate. The thin
samples naturally roll up into a cylinder unless supported on a
frame. (Inset) A flexible ``skin" applied to a finger.}
\vspace{0.2in}
\end{figure}

Closely related to creating THz metamaterial composites is the
seminal work on frequency selective surfaces
\cite{ulrich1,ulrich2,ulrich3}. More recently, there has been a
report on creating polyimide-based multilayer metallic photonic
band-gap (MPBG) structures at terahertz frequencies
\cite{gupta1997}. The MPBG structure served as a -35 dB notch filter
at $\sim$4.5 THz. The primary difference between \cite{gupta1997}
and the present publication is that (a) we are focusing on split
ring resonators and (b) our structures are easily peeled from the
substrate while in \cite{gupta1997}, a citric acid/hydrogen peroxide
etch was used to remove the substrate.

It is important to extend such work and develop MM on thin flexible
substrates that are considerably thinner than the lateral unit cell
dimensions in moving towards creating multi-layer non-planar
metamaterials such as cloaks, concentrators, or resonant absorbers
at terahertz frequencies.  In this letter, we demonstrate resonant
terahertz metamaterials on free standing low-loss polyimide
substrates as thin as 5.5 $\mu$m yielding robust large-area
metamaterials which are easily wrapped into cylinders with a radius
of a few millimeters.

The MM structures were fabricated by depositing a 200-nm-thick gold
with a 10-nm-thick adhesion layer of titanium on a polyimide
substrate. The liquid polyimide of PI-5878G (HD MicroSystems$^{TM}$)
was spin-coated on a 2 inch silicon wafer to form the substrate. In
this work, we fabricated our samples with two different thicknesses,
namely 5.5 $\mu$m and 11 $\mu$m. The thickness of the polyimide
substrate can be precisely controlled by adjusting the spin rate and
curing temperature. AZ5214e image reversal photoresist was patterned
using direct laser writing with a Heidelberg DWL 66 laser writer.
200 nm-thick Au/Ti was E-beam evaporated followed by rinsing in
acetone for several minutes. As a final step, the MM structures
patterned on the polyimide substrate were peeled off of the silicon
substrate. The as-fabricated 2 inch diameter samples show extreme
mechanical flexibility, as shown in Fig. 1. These samples can be
wrapped into cylinders with a radius of approximately three
millimeters. Harsh environmental tests have been conducted to test
the robustness of the free standing metamaterial samples by rinsing
them in organic solutions such as methanol and isopropanol, folding
and crumpling, and heating them up to 350$^{\circ}$C. No distortion
or cracking was observed and the electromagnetic resonances were
unaffected by this treatment.

\begin{table}
\caption{\label{tab:table1} Dimensions of each free standing metamaterial particles (all units in $\mu$m): a, lattice period; b, outer dimension; t, thickness of the polyimide substrate; w, line width; l, length of the gap; g, gap distance; for sample 3, the distance between inner and outer rings is 2$\mu$m. }
\begin{ruledtabular}
\begin{tabular}{ccccccc}
 Sample&$a$&$b$&$t$&$w$&$l$&$g$\\
\hline
1 & 50 & 36 & 5.5 & 4 & 8 & 2 \\
2 & 50 & 36 & 5.5 & 4 & 4 & 2 \\
3 & 50 & 36 & 5.5 & 4 & 4 & 4 \\
4 & 250 & 180 & 11.5 & 6 & 67 & 6 \\
\end{tabular}
\end{ruledtabular}
\end{table}

Terahertz time-domain spectroscopy (THz-TDS) was used to
characterize the metamaterial response. The transmission of the THz
electric field was measured for the sample and a reference, which in
the present case is simply air. The electric field spectral
amplitude and phase are calculated through Fourier transformation of
the time-domain pulses. Dividing the sample by the reference yields
the complex spectral transmission \cite{nishizawa05}. Prior to
measurement, the free standing MM samples were diced into
1cm$\times$1cm squares and mounted at normal incidence with respect
to the THz radiation. All of the measurements were performed at room
temperature in a dry ($<$ 0.1$\%$ humidity) air atmosphere. The THz
beam diameter was $\sim$3 mm, which was significantly smaller than
the sample dimensions.

\begin{figure}[h]
\includegraphics[scale=1.5]{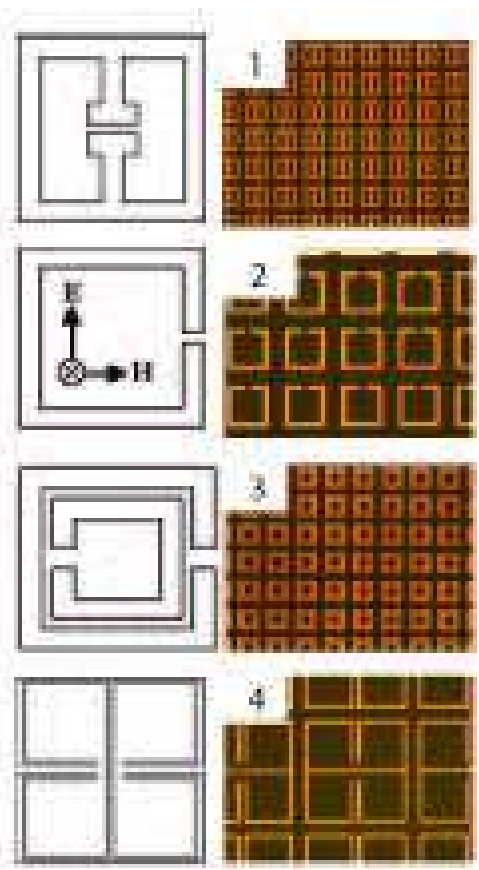}
\caption{\label{fig:Fig2} Designs and photographs of the polyimide
supported metamaterials. The corresponding dimensions are listed in
Table 1. The arrows labeled E and H show the electric and magnetic
field directions with respect to the orientation of all of the
SRRs.} \vspace{-0.2in}
\end{figure}

THz-TDS measurements were first carried out on a series of
substrates without metamaterials to characterize the complex
refractive index, $n = n_{r} + in_{i}$, of the polyimide samples.
Films with thicknesses of 5.5$\mu$m, 11 $\mu$m, 60 $\mu$m, and 160
$\mu$m were fabricated using spin-coating as described above. For
each of these films, a frequency independent refractive index
$n_{r}$ = 1.8 $\pm$ 0.05 was experimentally determined (from 0.2 -
2.5 THz). For all of the samples measured $n_{i}$ was also frequency
independent. However, there was considerable variability in the
magnitude of $n_{i}$ ranging from 0.026 up to 0.150. We attribute
this variation to imperfections in the polyimide substrates
resulting in scattering which yields an apparent increase in
$n_{i}$. The larger values were obtained from the thicker samples
consistent with this interpretation. For the thinnest films, the
value was typically $n_{i}$ = 0.04 and the field transmission was
greater than 0.95 from 0.2 to 2.5 THz. We note that at 1.0 THz
$n_{i}$ = 0.04 corresponds to a power absorption coefficient of
$\alpha = 2\omega n_{i}/c$ = 30 cm$^{-1}$. In short, polyimide
serves as a low-index low-loss and highly flexible substrate upon
which to fabricate THz metamaterials.

Numerous free-standing polyimide  /metamaterial samples were
fabricated and characterized using THz-TDS. For brevity, we focus on
four of these samples. Schematics of the metamaterial particles are
displayed in the left column of Fig. 2 while the right half of the
figure shows optical micrographs of portions of the arrays. The
samples include purely electric resonators (1 and 4) and canonical
split ring resonators (2 and 3).

\begin{figure}[h]
\includegraphics[scale=0.9]{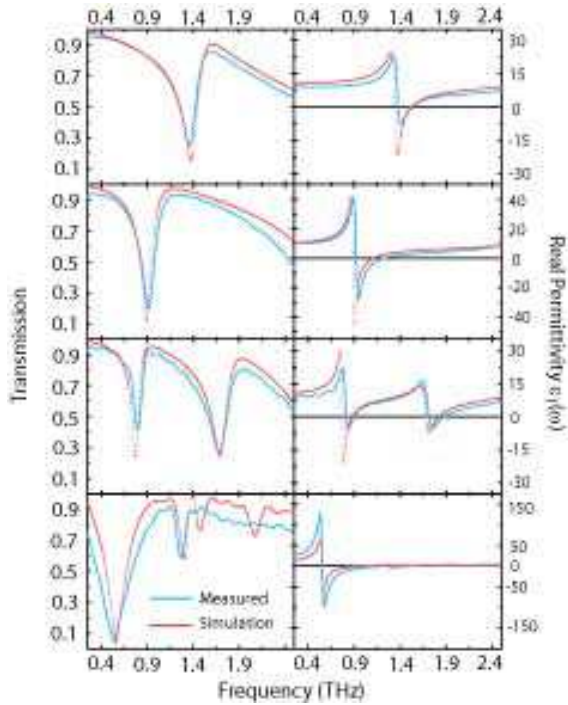}
\caption{\label{fig:Fig3} The left panels show the experimentally measured transmission for the corresponding samples in Figure 2. The blue line is experiment, and the red line is the simulated transmission. The right panels show the real part of the dielectric response. The blue lines are determined from the experimental data and the red lines are determined from the simulations.}
\end{figure}

The experimental and simulation results are displayed in Figure 3.
The left hand side shows (blue lines) the experimentally measured
field transmission as a function of frequency. The resonances are of
quality equal to those previously measured on rigid semiconducting
substrates \cite{padilla07}. The red lines are the results of
electromagnetic simulations using Microwave Studio where the
dimensions listed in Table I on were used for the SRR elements and
the experimentally measured refractive index for polyimide ($n = 1.8
+ i0.04$) was used. The agreement with experiment is quite good.

The right hand side of Figure 3 shows the results of extracting the
effective dielectric function ($\epsilon(\omega) = \epsilon_{1} +
i\epsilon_{2}$) for our thin films (only $\epsilon_{1}$ is plotted).
The determination of $\epsilon(\omega)$ followed the standard
approach described in previous publications
\cite{padilla06,chen06,chen08} with the exception that the effective
dielectric function of the entire MM/polyimide film was determined.
Hence, a thickness of 5.5 or 11$\mu$m, as appropriate, was used. In
addition, we determined the dielectric response from the simulated
transmission following the approach describe in reference
\cite{smith_ext_02,smith_ext_05}. The excellent agreement with the
experimental data attests to the high quality of the metamaterial
samples.

The use of spin-coating polyimide as a flexible substrate for
metamaterials offers numerous possibilities as functional
electromagnetic coatings including narrowband filters or absorbers.
Importantly, the polyimide thickness can be controlled with
sub-micron precision and, further, the total substrate thickness can
be substantially less than the lateral unit cell dimensions which
facilitates the incorporation into multilayer (potentially
heterogeneous) structures. The flexible substrate is of sufficiently
low loss to enable the construction of multilayer samples including,
potentially, perfect absorbers, negative index materials, and THz
electromagnetic cloaks. As described above, part of the substrate
loss arises from scattering due to imperfections. Such losses may be
further reduced with improvements in the spin-coating process.

In summary, flexible resonant terahertz metamaterials built on
ultrathin highly flexible polyimide substrates have been designed,
fabricated and measured. These results pave the way for creating
numerous multilayered non-planar electromagnetic composites.

We acknowledge partial support from the Los Alamos National
Laboratory LDRD program, DOD/Army Research Laboratory
W911NF-06-2-0040, NSF EECS 0802036, and DARPA HR0011-08-1-0044. H.
T. would like to thank Mr. George Seamans from Department of
Aerospace and Mechanical at Boston University for help on the
fabrication process. The authors would also like to thank the
Photonics Center at Boston University for all of the technical
support throughout the course of this research.


\end{document}